\pdfoutput=1
\documentclass[aps,prd,twocolumn,superscriptaddress,nofootinbib]{revtex4-2}
\usepackage{amsmath,amssymb,graphicx,hyperref,bm,booktabs}
\usepackage{xcolor}
\usepackage[normalem]{ulem}
\hypersetup{colorlinks=true,linkcolor=blue,citecolor=blue,urlcolor=blue}

\begin{document}

\title{DESI and Dynamical Dark Energy from Extended Pre-geometric Gravity}

\author{Andrea Addazi}
\email{addazi@scu.edu.cn}
\affiliation{School of Physics and Astronomy, Anqing Normal University, Anqing 246133, People's Republic of China;}
\affiliation{Institute of Astronomy and Astrophysics, School of Mathematics and Physics, Anqing Normal University, Anqing 246133, China;}
\affiliation{Laboratori Nazionali di Frascati INFN, Frascati 00044, Italy;}

\author{Yermek Aldabergenov}
\affiliation{CAS Key Laboratory for Researches in Galaxies and Cosmology/Department of Astronomy, School of Astronomy and Space Science, University of Science and Technology of China, Hefei, Anhui 230026, China;}
\affiliation{Department of theoretical and nuclear physics, Al-Farabi Kazakh National University, Almaty 050040, Kazakhstan;}

\author{Daulet Berkimbayev}
\affiliation{Department of theoretical and nuclear physics, Al-Farabi Kazakh National University, Almaty 050040, Kazakhstan;}

\author{Yifu Cai}
\affiliation{Deep Space Exploration Laboratory/School of Physical Sciences,
University of Science and Technology of China, Hefei, Anhui 230026, China;}
\affiliation{CAS Key Laboratory for Researches in Galaxies and Cosmology/Department of Astronomy, School of Astronomy and Space Science, University of Science and Technology of China, Hefei, Anhui 230026, China;}

\author{Salvatore Capozziello}
\affiliation{Dipartimento di Fisica  ``E.\ Pancini'', Università degli Studi di Napoli ``Federico II'', Compl.\ Univ.\ di Monte S.\ Angelo, Edificio G, Via Cinthia, I-80126, Napoli, Italy;}
\affiliation{Istituto Nazionale di Fisica Nucleare (INFN), Sez.\ di Napoli, Compl.\ Univ.\ di Monte S.\ Angelo, Edificio G, Via	Cinthia, I-80126, Napoli, Italy;}
\affiliation{Scuola Superiore Meridionale, Via Mezzocannone 4, I-80134, Napoli, Italy;}

\author{Giuseppe Meluccio}
\affiliation{Scuola Superiore Meridionale, Via Mezzocannone 4, I-80134, Napoli, Italy;}
\affiliation{Istituto Nazionale di Fisica Nucleare (INFN), Sez.\ di Napoli, Compl.\ Univ.\ di Monte S.\ Angelo, Edificio G, Via	Cinthia, I-80126, Napoli, Italy;}
\affiliation{Department of Physics, Loyola Marymount University,  90045, Los Angeles, California, USA.}

\date{\today}

\begin{abstract}
We consider the simplest quadratic extension of MacDowell--Mansouri pre-geometric gravity that preserves the topological pre-volume-form symmetry. Upon symmetry breaking, the theory reduces to $(\mathrm{Lovelock})^2$ gravity and admits a Galileon-like Horndeski scalar-tensor representation. The gravitational Higgs mechanism relates the Gauss--Bonnet coupling to the bare cosmological constant, while the quadratic correction elevates the gravitational $\theta$-angle to a dynamical gravi-axion whose ultra-light mass scale can drive late-time acceleration. We confront the Jordan-frame background with DESI DR2 baryon-acoustic-oscillation measurements and an auxiliary six-point compressed growth diagnostic. 
 A transient dark-energy response localized near $H\simeq m_b$ ($m_b$ the effective gravi-axion mass) reproduces the fitted expansion and its BAO-distance predictions and becomes negligible at high redshift. For the full linear calculation, we combine a smooth positive-density representative of this response class with explicitly imposed GR-like tensor and local sectors, together with the parameterized post-Friedmann prescription. The resulting gravi-axion benchmark exhibits a transient effective phantom crossing and admits a regular full linear CAMB evolution. Its separate likelihood contributions are promising relative to those obtained from the flat-$\Lambda$CDM baseline for the same data set.
  These results establish the dynamical gravi-axion framework, dual to MacDowell--Mansouri pre-geometric gravity, as a viable effective quantum-gravity inspired description of late-time cosmology, motivating further investigation into its fundamental origin and perturbative consistency.
\end{abstract}

\maketitle


\vspace{0.1cm}

{\it Introduction}. Gauge-theoretic approaches to gravity, notably those of MacDowell--Mansouri (MM) \cite{macdowell:unified} and Wilczek (W) \cite{wilczek:gauge}, offer a compelling emergence scenario: gravity arises as the broken phase of a Yang--Mills theory with de Sitter gauge group $SO(1,4)$ on a manifold that initially lacks a metric. A Higgs field $\phi^A$ spontaneously breaks the symmetry to the Lorentz group $SO(1,3)$, dynamically generating the vierbein, the spin connection, the Einstein--Hilbert action and the cosmological constant (CC) term. These minimal models are remarkably rigid \cite{Addazi:2024rzo,Addazi:2025vbw,Meluccio:2025uyo,Addazi:2025gcl}, admitting only two consistent Lagrangian densities under standard assumptions.

In this letter, we extend the MM model with a quadratic correction preserving the topological current $J$, which is related to the pre-geometric volume density. In MM, symmetry breaking forces a Gauss--Bonnet (GB) coupling to be related to the CC as $\alpha_{\text{GB}} \sim M_\textup{P}^2/\Lambda \sim 10^{120}$, which coincides with the de Sitter entropy of the observable Universe \cite{Addazi:2026kam,HNPGG,HNPGG2,Addazi:2020axm,Addazi:2020wnc,Addazi:2020mnm}. While the GB term is topological in four dimensions for linear gravity, quadratic terms render it dynamically significant. The theory can be recast as a Galileon-like Horndeski scalar-tensor model where the canonical scalar field acquires an ultra-light mass, naturally behaving as dynamical dark energy, and the large GB coefficient can be interpreted as a gravitational $\theta$-angle promoted to a gravi-axion.

We compare the resulting Jordan-frame expansion history with the DESI DR2 BAO measurements \cite{DESI:DR2BAO} and with an auxiliary compressed growth-rate diagnostic using the six redshift bins of the DESI DR1 full-shape analysis \cite{DESI:DR1FS}. The background admits a good low-redshift fit, but an exact perturbative implementation of the unmodified covariant action reveals a tensor-gradient instability and fails the multimessenger constraint on the propagation speed of gravitational waves \cite{GW170817GRB}. 
Such instabilities arise only in the regime where the gravi-axion mass---which corresponds to an exponential term in the potential of $\varphi$ field---vanishes. This regime is regarded as unphysical.
We therefore do not use the unmodified action for linear observables. Instead, we formulate the viable late-time sector as an effective gravi-axion framework, in which we supplement the theory with a natural mass term for the gravi-axion, interpreted as arising from gravitational instantons. This effective mass localizes a vacuum-like response around the release epoch $H\simeq m_b$, preserves the fitted Jordan-frame expansion over the observational interval, and becomes negligible at high redshift. For the full Boltzmann calculation, the tensor and local gravitational sectors are taken to be GR-like and the effective dark-energy perturbations are evolved with the PPF prescription \cite{Fang:2008sn} in CAMB \cite{Lewis:1999bs}, allowing a controlled transient crossing of $w_{\rm eff}=-1$. The final BAO and growth likelihoods are statistically close to the common-data $\Lambda$CDM baseline, while the full CMB, lensing, transfer-function and matter-power calculations remain regular. The gravi-axion mass may be generated by gravitational instantons and becomes comparable to the Hubble rate at a critical redshift $z_c \sim 0.5\div 3.0$. A related late-time nonequilibrium-axion construction has shown how an effective axion response can modify low-redshift cosmology while recovering $\Lambda$CDM at earlier times\cite{Montani:2026Axion}. In the present setting the gravi-axion instead acts as a vacuum-like dark-energy degree of freedom. The gravi-axion response organizes the physical content of the effective late-time sector, while tensor luminality and the PPF perturbation are dynamically accounted for by the gravi-axion evolution.

\vspace{0.2cm}

{\it Gravitation from a Spontaneously Broken Gauge Theory}.
We consider a gauge theory over a four-dimensional manifold, \emph{prior} to the introduction of a spacetime metric. The gauge group is taken to be the de Sitter \(SO(1,4)\) one, with gauge potentials \(A_{\mu}^{AB}\) and field strengths \(F_{\mu\nu}^{AB}\). The theory is required to be generally covariant, yet no metric or tetrad is assumed \textit{a priori}. The only fixed structure on each tangent space is an internal metric of signature \((-,+,+,+,+)\), generalizing the Minkowski metric \(\eta\).

To dynamically generate a Riemannian geometry, we introduce a scalar field \(\phi^A\) in the internal vector representation. A spontaneous symmetry breaking (SSB) in its vacuum expectation value (VEV) $v$ reduces the gauge group to the Lorentz group \(SO(1,3)\) \cite{wilczek:gauge}. Below the symmetry-breaking scale, the effective theory describes gravity emerging from the interactions of the pre-geometric fields \(A_{\mu}^{AB}\) and \(\phi^A\).

General covariance demands that Lagrangian densities be scalar densities of weight \(-1\). In the absence of an inverse metric, the only intrinsic contravariant object on the manifold is the Levi-Civita symbol \(\epsilon^{\mu\nu\rho\sigma}\), a tensor density of weight \(-1\). Using it, two generally covariant Lagrangian densities can be constructed:

\vspace{0.1cm}

1. the MacDowell--Mansouri (MM) Lagrangian \cite{macdowell:unified}:
\begin{equation}
\label{MM}
\mathcal{O}_{\text{MM}} = k_{\text{MM}} \, \epsilon_{ABCDE} \, \epsilon^{\mu\nu\rho\sigma} \, F_{\mu\nu}^{AB} F_{\rho\sigma}^{CD} \, \phi^{E},
\end{equation}

\vspace{0.1cm}

2. the Wilczek (W) Lagrangian \cite{wilczek:gauge}:
\begin{equation}
\label{W}
\mathcal{O}_{\text{W}} = k_{\text{W}} \, \epsilon_{ABCDE} \, \epsilon^{\mu\nu\rho\sigma} \, F_{\mu\nu}^{AB} \, \nabla_{\!\rho}\phi^{C} \, \nabla_{\!\sigma}\phi^{D} \, \phi^{E},
\end{equation}
where the gauge-covariant derivative acts as
\begin{equation}
\nabla_{\mu}\phi^{A} = \partial_{\mu}\phi^{A} + A_{B\mu}^{A} \phi^{B},
\qquad
A_{B\mu}^{A} \equiv \eta_{BC} A_{\mu}^{AC}.
\end{equation}
Here uppercase Latin indices run from 1 to 5, Greek indices from 1 to 4, and the couplings have dimensions \([k_{\text{MM}}] = [\phi]^{-1}\), \([k_{\text{W}}] = [\phi]^{-3}\) ($\eta_{BC}$ is the internal space flat metric of $SO(1,4)$) \cite{Addazi:2024rzo}.

We first examine the low-energy effective theory after the SSB, where the internal direction is fixed by \(\langle\phi^{A}\rangle = v \delta^{A}_{5}\).  
The gauge potentials then split into
\[
A_{\mu}^{a5} \quad\text{(four components)},\qquad
A_{\mu}^{ab} \quad\text{(six components)},
\]
with lowercase Latin indices \(a,b = 1,\dots,4\). Identifying
\begin{equation}\label{eq:id}
e_{\mu}^{a} \equiv m^{-1} A_{\mu}^{a5},\qquad 
\omega_{\mu}^{ab} \equiv A_{\mu}^{ab},
\end{equation}
where \(e_{\mu}^{a}\) are the tetrads, \(\omega_{\mu}^{ab}\) the spin connection and \(m\) a mass parameter, the broken-phase Lagrangians become exactly those of general relativity.

To understand how this happens, let us discuss the case of the MM theory in more detail. Substituting the vacuum expectation value of \(\phi^{A}\) into \(\mathcal{O}_{\text{MM}}\) yields
\begin{align}
\label{SSBMM}
\mathcal{O}_{\text{MM}} \;\xrightarrow{\text{SSB}}\;
\ 16 k_{\text{MM}} v m^{2} \, e \, e_{a}^{\mu} e_{b}^{\nu} R_{\mu\nu}^{ab}
- 96 k_{\text{MM}} v m^{4} \, e
\nonumber\\
- 4 k_{\text{MM}} v \, e \, \mathcal{G},
\end{align}
where \(e = \det e_{\mu}^{a}\), \(R_{\mu\nu}^{ab}\) is the curvature two-form constructed from \(\omega_{\mu}^{ab}\) (i.e., the Riemann tensor) and \(\mathcal{G}\) is the GB term. The three terms above correspond, respectively, to the Einstein--Hilbert action, the cosmological constant term and a topological invariant. Matching them to standard gravity gives rise to the emergent (reduced) Planck mass and the emergent CC respectively as
\begin{equation}
\label{KE1}
M_{\text{P}}^{2} \equiv  32 k_{\text{MM}} v m^{2},\qquad \lambda \equiv 3 m^{2} = \frac{3 M_{\text{P}}^{2}}{32 k_{\text{MM}} v}.
\end{equation}
For \(k_{\text{MM}} \sim \mathcal{O}(1)\) and the observed value \(M_{\text{P}}^{2} \sim 10^{37}\ \text{GeV}^{2}\), the small value \(\Lambda \sim 10^{-84}\ \text{GeV}^{2}\) follows from a large VEV \(v \sim 10^{119\div 120}\ [\phi]\) \cite{Addazi:2024rzo}.

The W model relies instead on the role of the gauge-covariant derivative of the scalar field $\phi^A$, which after the SSB reduces to
\begin{equation}
\nabla_{\mu}\phi^{A} \;\xrightarrow{\text{SSB}}\; v A_{5\mu}^{A} =  v A_{\mu}^{a5}.
\end{equation}
Using the same identifications \eqref{eq:id} for \(e_{\mu}^{a}\) and \(\omega_{\mu}^{ab}\), this time we obtain
\begin{equation}
\label{SSBW}
\mathcal{O}_{\text{W}} \;\xrightarrow{\text{SSB}}\;
-4 k_{\text{W}} v^{3} m^{2} \, e \, e_{a}^{\mu} e_{b}^{\nu} R_{\mu\nu}^{ab}
+ 48 k_{\text{W}} v^{3} m^{4} \, e.
\end{equation}
This produces the Einstein--Hilbert term with a cosmological constant, with no GB contribution. In this case, the emergent gravitational scales are
\begin{equation}
M_{\text{P}}^{2} \equiv -8 k_{\text{W}} v^{3} m^{2},
\qquad
\Lambda \equiv  6 m^{2} = - \frac{3 M_{\text{P}}^{2}}{4 k_{\text{W}} v^{3}}.
\end{equation}
Taking \(|k_{\text{W}}| \sim \mathcal{O}(1)\) (notice that $k_W<0$ is the consistent choice) and the observed value of \(M_{\text{P}}^{2}\), the measured \(\Lambda\) corresponds to \(v \sim 10^{40}\ [\phi]\) \cite{Addazi:2024rzo}.

\vspace{0.2cm}

{\it On the Rigidity of Pre-geometric Gravity and the Emergent Gauss--Bonnet Coupling}. The framework under analysis is defined on a four-dimensional spacetime equipped solely with the Levi-Civita symbol $\epsilon^{\mu\nu\rho\sigma}$, while tangent spaces carry both $\eta_{AB}$ and $\epsilon_{ABCDE}$. No dynamical spacetime metric is assumed a priori, hence the pre-geometric spacetime is only axial-topological \cite{Addazi:2024rzo,Addazi:2025vbw,Addazi:2025gcl}. Under the hypotheses (i) four-dimensionality, (ii) general covariance, (iii) recovery of the Einstein–Hilbert action and the equivalence principle in the broken phase, and (iv) minimal pre-geometric field content (only $A_\mu^{AB}$ and $\phi^A$), the possible Lagrangians are uniquely determined \cite{Addazi:2024rzo}. Covariant derivatives of the field strength are excluded, leaving only three combinations that can supply the required four covariant indices: two field strengths $F_{\mu\nu}^{AB}$, one $F_{\mu\nu}^{AB}$ with two $\nabla_\mu\phi^A$, or four $\nabla_\mu\phi^A$. The first two do indeed reproduce $\mathcal{L}_\textup{EH}$, while the third yields only a cosmological constant. The internal Levi-Civita symbol forces the gauge group to be $SO(1,4)$ (for a positive CC). Using a single Levi-Civita symbol gives precisely the MM and W Lagrangians; multiple symbols either violate condition (iii) or reduce to these two.

From these building blocks, we can thus construct only \emph{extended} pre-geometric theories such as\footnote{Note that imposing a pre-geometric unimodular constraint $J=1$ (normalized in natural units) would simplify 
the Lagrangian density to $\mathcal{L}_{f\textup{MM}}=f(\mathcal{O}_\textup{MM})$. In this work, however, we will not impose this constraint -- which in principle can always be implemented via a Lagrange multiplier.}
\begin{equation}
\mathcal{L}_{f\textup{MM}}={\frac{1}{2}}J f( J^{-1}\mathcal{O}_\textup{MM})\,, \qquad 
\mathcal{L}_{f\textup{W}}={\frac{1}{2}}J f( J^{-1}\mathcal{O}_\textup{W})\,,
\end{equation}
where
\begin{equation}
\label{J}
J=\epsilon_{ABCDE}\epsilon^{\mu\nu\rho\sigma}\nabla_\mu\phi^A\nabla_\nu\phi^B\nabla_\rho\phi^C\nabla_\sigma\phi^D\phi^E.
\end{equation}
After the SSB, we find schematically that
\begin{equation}
\begin{aligned}
J&\xrightarrow{SSB} e\,,\qquad
\mathcal{L}_{f\textup{MM}}\xrightarrow{SSB} {\frac{1}{2}}e f(L)\,,\\
\mathcal{L}_{f\textup{W}}&\xrightarrow{SSB} {\frac{1}{2}}e f(R-2\lambda)\,.
\end{aligned}
\end{equation}
where $L\equiv R+\alpha_{\text{GB}}\mathcal{G}-2\lambda$. The $f(L)$ theory will be our focus in this work.~\footnote{Modified gravity of the $f(L)$ type was studied in \cite{Bueno:2016dol} in general dimensions. Its quadratic realization, $f(L)\sim L+L^2$, was studied in the context of inflation in \cite{Addazi:2025agg} as a one-parameter extension of the Starobinsky model.}

From the MM case, symmetry breaking yields the parametric relation
\begin{equation}
{{\alpha^{(0)}_{\text{GB}} = -8k_{\text{MM}}v = -\frac{3M_{\text{P}}^2}{4\lambda} \sim -10^{120}}} \qquad (\lambda\sim\Lambda)\,,
\label{eq:alpha-gb-large}
\end{equation}

{ where $^{(0)}$ means that this relation is valid for a "bare" Gauss-Bonnet coupling un-affected by any quantum and renormalization corrections\footnote{{Another possibility is to include in the Higgs potential a pre-geometric term of the form $\vert J\vert$, as defined in Eq.~\ref{J}, multiplied by a constant $f_{\mathrm{CC}}\vert J\vert$, where $f_{\mathrm{CC}}$ is a free parameter that may also take either sign. Following spontaneous symmetry breaking (SSB), this term contributes to the cosmological constant as $\delta\lambda\, e$, with $\delta\lambda = f_{\mathrm{CC}} m^{4} v^{5}$. The full bare cosmological constant then becomes $\lambda + \delta\lambda$, while only the $\lambda$ component remains linked to the Gauss--Bonnet coupling through Eq.~\ref{eq:alpha-gb-large}. In this scenario, a theory based on the Anti-de Sitter gauge group SO(2,3) — in addition to the de Sitter group SO(1,4) considered here — becomes viable, provided that the total $\lambda + \delta\lambda$ is positive. In that case, the sign of the bare Gauss--Bonnet angle in Eq.~\ref{eq:alpha-gb-large} would flip to positive as well. }}. }
Such a coupling is 
an enormous number comparable to the de Sitter entropy of the observable Universe:
\begin{equation}
{|\alpha_{\text{GB}}^{(0)}| \sim S_\textup{dS} \sim \frac{R_{H}^2}{\ell_{\text{P}}^2}\equiv N\,,}
\end{equation}
with $R_{H}$ being the Hubble radius, $\ell_{\text{P}}$ the Planck length and $N$ the number of holographic qubits of information encoded on the cosmological horizon. Therefore, the pre-geometric Higgs field $\phi^A$ also acts as an \textit{information field}, with the largeness of $\alpha_{\text{GB}}$ arising from a see-saw mechanism between the large VEV $v$ and the small CC.\footnote{Equivalently, one has that
\begin{equation}\label{alpha_vs_CC}
{|\alpha_{\text{GB}}^{(0)}| \sim \frac{1}{\alpha_G(\Lambda)}}\,,\quad \alpha_G(\Lambda)=\frac{\Lambda}{M_{\text{P}}^2}\,,
\end{equation}
suggesting an electric-magnetic duality reminiscent of Dirac~\cite{Dirac}, Wu--Yang~\cite{WuYang}, Montonen--Olive \cite{Montonen:1977sn} and Seiberg--Witten~\cite{SW} monopoles.} The quantization of the GB charge then naturally relates $\alpha_\textup{GB}$ and the gravitational coupling $\alpha_G$ as
\begin{equation}
{|\alpha_{\text{GB}}^{(0)}|\,\alpha_G(\Lambda)\sim 1 \;\Longrightarrow\; S_\textup{dS}\sim|\alpha_{\text{GB}}^{(0)}|\sim N\,.}
\end{equation}

In four dimensions, the GB term is topological for a linear Lagrangian, contributing only as a boundary term. For a closed manifold, indeed,
\begin{equation}
{\alpha_{\text{GB}}^{(0)}\int e\,\mathcal{G}\,d^4x = 32\pi^2\alpha_{\text{GB}}^{(0)}\,\chi(M)\, , }
\end{equation}
with $\chi(M)$ being the Euler characteristic. Hence, $\alpha_{\text{GB}}$ acts as a gravitational $\theta$-angle, i.e.\ {$\theta=32\pi^2\alpha_\textup{GB}^{(0)}$}. As remarked in Ref.\ \cite{Addazi:2026kam}, this also forces the pre-geometric Higgs potential to be periodic, subject to the topological pre-geometric current conservation, and the CC $\lambda$ to be quantized as $\lambda \sim M_\textup{P}^2/k$, 
with $k=1,\dots,N,\dots$. 

{Concerning the sign of the combination $\alpha_{GB}^{(0)}\,\chi$, there is, in general Gauss--Bonnet theories, an ambiguity not unlike that of $\theta$-term gauge theories. For instance, the overall sign can be influenced by the topological properties of the background---whether of classical origin (e.g., topological defects) or quantum nature (e.g., instantons). In the case of closed manifolds, where $\chi = 2 - 2g$ with $g$ the genus, the value of $\chi$ can be altered by the inclusion of gravitational instantons or realistic topological defects such as cosmic strings. In principle, the resulting sign should be obtained by summing over all relevant semiclassical saddle-point solutions.

From a low-energy effective perspective relevant to cosmology, one may consider two possible branches for the combined term $\alpha_{GB}^{(0)}\,\chi$: one with an overall positive sign and one with a negative sign. In the presence of the Einstein--Hilbert action plus the Gauss--Bonnet term alone, this ambiguity is immaterial, as it does not directly affect any physical observables.

However, as we shall see, once higher-order corrections are included, $\alpha_{GB}$ becomes a dynamical field, and the Gauss--Bonnet density enters into the field equations. Consequently, the sign ambiguity acquires physical significance and must be resolved by a choice. The branch that appears to be compatible with observational data is the one with a positive overall sign. Therefore, in what follows, we take the effective coupling to be $\alpha_{GB} = \alpha_{GB}^{(0)}\,\chi$, whose sign is determined by underlying topological data.

Indeed, considering a slowly varying field around the topological saddle solution, the sign is expected to be determined by topological data as well, since small variation of the field around the standard solution. The positive branch is selected on the basis of what we may call the "entropic razor principle": only large positive contributions to the de Sitter entropy are considered reasonable and acceptable, whereas large negative ones must be excluded.}

The main point of interest here is that, for a generic $f(L)$ theory, the GB term is no longer topological and therefore enters the equations of motion dynamically. Consequently, the enormous value of $\alpha_{\text{GB}}$ becomes physically significant. This is dual to promoting the gravitational $\theta$-angle to a dynamical super-light axion-like-particle, as will be discussed in the next sections.
Indeed, it can be shown that including a topological $\Theta$-term alongside the (extended) MM term yields an elegant linear relation between the Pontryagin $\theta$-term and the Gauss--Bonnet term, as a unique prediction of the theory. Consequently, the dynamical $\theta$-angle coincides precisely with the gravi-axion arising from the potential term, albeit equipped with a non-minimal Galileon- or Horndeski-like kinetic term.

\vspace{0.2cm}

{\it Emergent $f(L)$ Model}. The emergent modified gravitational theory from the extended MM model corresponds to four-dimensional $f({\rm Lovelock})$ gravity, defined by the Lagrangian density (in Planck units, $M_\textup{P}=1$)
\begin{equation}
\label{Lov1}
\mathcal{L}=\frac{1}{2}e~f(L),\qquad L=-2\lambda+R+\alpha_\text{GB}\mathcal{G},
\end{equation}
where $L$ represents a 4D Lovelock-type combination containing a cosmological constant term $\lambda$ and a GB coupling parameter $\alpha_{\rm GB}$. An alternative form of this Lagrangian is
\begin{equation}
\label{Lov2}
e^{-1}\mathcal{L}=\frac{1}{2}\big[f'(Z)L-f'(Z)Z+f(Z)\big],
\end{equation}
in which $Z$ is an auxiliary scalar. Variation with respect to $Z$ recovers $Z=L$, provided that $f''(Z)\neq 0$, thereby returning the original Lagrangian \eqref{Lov1}. Notably, only one scalar degree of freedom is required for a scalar-tensor description of this theory, in contrast to generic $f(R,\mathcal{G})$ models that generally involve two scalars and often contain ghost instabilities as well \cite{DeFelice:2010hg} (see also Refs.\ \cite{Bamba:2010wfw,Elizalde:2010jx,Atazadeh:2013cz,Nojiri:2021mxf}).

It is instructive to examine the Einstein-frame formulation of the theory in order to make the canonical scalar kinetic term and the second-order Horndeski structure explicit. A Weyl rescaling of the metric, $g_{\mu\nu}\rightarrow \frac{1}{f'}g_{\mu\nu}$, transforms the Lagrangian \eqref{Lov2} from the Jordan to the Einstein frame (up to boundary terms), yielding \cite{Addazi:2025agg}
\begin{align}
\begin{aligned}
\label{LL}
\frac{\mathcal{L}}{\sqrt{-g}} &=\frac{1}{2}R-\frac{3}{4f'^{2}}\partial f' \partial f'-\frac{1}{2f'}\Big(Z+2\lambda-\frac{Z}{f'}\Big)\\
&\quad + \frac{\alpha_\text{GB}}{2}\Big[f'\mathcal{G}+\frac{4}{f'}G^{\mu\nu}\partial_{\mu} f' \partial_{\nu} f'\\
&\qquad\quad -\frac{3}{f'^2}\partial f' \partial f'\Box f'+\frac{3}{f'^{3}}(\partial f' \partial f')^{2}\Big].
\end{aligned}
\end{align}
Here $\Box\equiv\nabla_\mu\nabla^\mu$, with $\nabla_\mu$ being the covariant derivative, and $G_{\mu\nu}=R_{\mu\nu}-\frac{1}{2}g_{\mu\nu}R$ is the Einstein tensor.

 The derivative $f'$ is then expressed in terms of a canonical scalar field $\varphi$ through the redefinition $f'\equiv e^{\sqrt{\frac{2}{3}}\varphi}$. Substituting this relation into Eq.\ \eqref{LL} yields the Einstein-frame Lagrangian in its final form:
\begin{align}
\begin{aligned}
\label{L_canonical}
&\sqrt{-g}^{-1}\mathcal{L}=\frac{1}{2}R-\frac{1}{2}\partial \varphi \partial \varphi -V(\varphi)\\
&\quad +\frac{\alpha_\text{GB}}{2} e^{\sqrt{\frac{2}{3}}\varphi}\Big(\mathcal{G}+\frac{8}{3}G^{\mu\nu}\partial_{\mu}\varphi\partial_{\nu}\varphi-\sqrt{\frac{8}{3}}\partial \varphi \partial \varphi \Box \varphi\Big),
\end{aligned}
\end{align}
with the scalar potential
\begin{equation}
    V(\varphi)=\tfrac{1}{2}e^{-\sqrt{\frac{2}{3}}\varphi}\big[Z(\varphi)+2\lambda-e^{-\sqrt{\frac{2}{3}}\varphi}f(Z(\varphi))\big].
\end{equation}
The Lagrangian \eqref{L_canonical} belongs to the Horndeski class \cite{Horndeski:1974wa}, also known as generalized galileons \cite{Deffayet:2011gz}, which constitutes the most general scalar-tensor theory with second-order equations of motion (see, e.g., Ref.\ \cite{Kobayashi:2019hrl} for a review).

Among the Horndeski operators, the cubic Galileon term $(\partial\varphi)^2\Box\varphi$ and the non-minimal derivative coupling $G^{\mu\nu}\partial_\mu\varphi\partial_\nu\varphi$ play a particularly important role in our setup. These are precisely the terms that naturally emerge from the extended MacDowell--Mansouri formulation of gravity, and they are exactly the ones responsible for activating the Vainshtein mechanism on Solar-System scales~\cite{Vainshtein:1972, Nicolis:2008in,Babichev:2013, Koyama:2013}. Their derivative self-interactions become non-negligible in high-density environments, leading to nonlinear field equations that strongly suppress the propagation of the fifth force within the Vainshtein radius, thereby ensuring compatibility with local gravity constraints \cite{Kase:2013uja}. For example, consider the term $(\partial\varphi)^2\Box\varphi$ which has the coefficient $\alpha_{\rm GB}e^{\sqrt{2/3}\varphi}$ in the Lagrangian \eqref{L_canonical}. For the selected late-time benchmark, the present scalaron value is sub-Planckian, $\varphi_0<1$ (more precisely, $\varphi_0=0.56222$ in reduced Planck units, $M_\textup{P}=1$), as obtained from the numerical solution discussed below. Accordingly, $e^{\sqrt{2/3}\varphi_0}=1.5826$ remains an order-unity factor and does not alter the parametric estimate; for the Vainshtein mechanism to operate, we therefore require $\alpha_{\rm GB}\sim1/\Lambda$ \cite{Nicolis:2008in}, where $\Lambda$ is the cosmological constant. Remarkably, this is exactly what our framework requires, as shown by Eq.~\eqref{alpha_vs_CC}.

\vspace{0.2cm}

{\it The Quadratic Case}. To illustrate the features of this framework, we choose a specific function $f(L)$ expanded up to quadratic order: $f(L)=L+\frac{L^2}{6M^{2}}$. For this choice, the relations following from Eq.\ \eqref{Lov2} become
\begin{gather}
\begin{gathered}
f(Z)=Z+\frac{Z^{2}}{6M^{2}},\qquad f'(Z)=e^{\sqrt{\frac{2}{3}}\varphi}=1+\frac{Z}{3M^{2}},\\
\Rightarrow~~Z(\varphi)=3M^{2}\Big(e^{\sqrt{\frac{2}{3}}\varphi}-1\Big),
\end{gathered}
\end{gather}
leading to the scalar potential
$V=V_{1}+V_{2}$,
with $V_{1}(\varphi)=\tfrac{3}{4}M^{2}\Big(1-e^{-\sqrt{\frac{2}{3}}\varphi}\Big)^{2}$
and $V_{2}(\varphi)=\lambda e^{-\sqrt{\frac{2}{3}}\varphi}$,
which matches the Starobinsky $R^2$ inflation potential augmented by a term proportional to $\lambda$. A positive $\lambda$ can shift the vacuum from Minkowski to de Sitter, potentially describing dark energy when sufficiently small. Writing $y\equiv e^{-\sqrt{2/3}\varphi}$, the potential  is minimized for $\langle y\rangle=1-\frac{2\lambda}{3M^2}=1-2\frac{m^2}{M^2}$, where we made use of the pre-geometric relation $\lambda=3m^2$. Therefore, the existence of a minimum requires that $m^2/M^2<1/2$, and the cosmological constant is corrected to 
\begin{equation}\label{true_CC}
    \Lambda\equiv\langle V\rangle=\lambda\Big(1-\frac{\lambda}{3M^2}\Big)=3m^2\Big(1-\frac{m^2}{M^2}\Big)~.
\end{equation}
Since $m^2/M^2<1/2$, we have that $3m^2/2<\Lambda\lesssim 3m^2$. Around the minimum of Eq.\ \eqref{true_CC}, the mass of the scalaron $\varphi$ is
$m_\varphi=M\Big(1-2\frac{m^2}{M^2}\Big)$. In other cases, the dynamics will coincide with a slow runaway.
{We emphasize that in this model, $M$ is treated as an independent parameter, distinct from $m$, and will be fixed by phenomenological fitting. However, barring fine-tuned cancellations, $M$ is naturally of the same order as $m$, since it is inversely proportional to the absolute value of the Gauss--Bonnet coupling, $M^{2} \propto |\alpha_{GB}|^{-1}$.}



We now present the equations of motion derived from the Lagrangian \eqref{L_canonical}. The Klein--Gordon equation and the Einstein equations read respectively
\vspace{0.2cm}
\onecolumngrid
\begin{align}
\begin{split}\label{KG}
    &\Box \varphi- V_{,\varphi}-\tfrac{1}{8}\xi_{,\varphi}\Big(\mathcal{G}+\tfrac{8}{3}G^{\mu\nu}\partial_\mu\varphi\partial_\nu\varphi-\sqrt{\tfrac{8}{3}}\partial\varphi\partial\varphi\Box\varphi\Big)\\
    &\quad +\tfrac{1}{\sqrt{6}}\partial_\mu\xi\Big(\sqrt{\tfrac{8}{3}}G^{\mu\nu}\partial_\nu\varphi-\partial^\mu\varphi\Box\varphi+2\nabla^\mu\nabla^\nu\varphi\partial_\nu\varphi\Big)\\
    &\quad +\tfrac{1}{2\sqrt{6}}\Box\xi\partial\varphi\partial\varphi\\
    &\quad +\tfrac{1}{\sqrt{6}}\xi\Big(\sqrt{\tfrac{8}{3}}G^{\mu\nu}\nabla_\mu\partial_\nu\varphi+R^{\mu\nu}\partial_\mu\varphi\partial_\nu\varphi-\Box\varphi\Box\varphi+\nabla^\mu\nabla^\nu\varphi\nabla_\mu\nabla_\nu\varphi\Big)=0,
\end{split}\\[10pt]
\begin{split}\label{EFE}
    &(1+\Box\xi)G_{\mu\nu}-\partial_\mu\varphi\partial_\nu\varphi+\tfrac{1}{2}g_{\mu\nu}(\partial\varphi\partial\varphi+2V)\\
    &\quad +\tfrac{1}{2}\nabla_\mu\nabla_\nu\xi\,R+\nabla^\rho\nabla^\sigma\xi\,R_{\mu\rho\sigma\nu}+g_{\mu\nu}\nabla^\rho\nabla^\sigma\xi\,R_{\rho\sigma}\\
    &\quad -\nabla^\rho\nabla_\mu\xi\,R_{\rho\nu}-\nabla^\rho\nabla_\nu\xi\,R_{\rho\mu}\\
    &\quad +\tfrac{1}{3}\nabla_\lambda\nabla_\mu(\xi\partial_\nu\varphi\partial^\lambda\varphi)+\tfrac{1}{3}\nabla_\lambda\nabla_\nu(\xi\partial_\mu\varphi\partial^\lambda\varphi)\\
    &\quad -\tfrac{1}{3}\nabla_\mu\nabla_\nu(\xi\partial\varphi\partial\varphi)-\tfrac{1}{3}\Box(\xi\partial_\mu\varphi\partial_\nu\varphi)\\
    &\quad +\tfrac{1}{3}g_{\mu\nu}\Box(\xi\partial\varphi\partial\varphi)-\tfrac{1}{3}g_{\mu\nu}\nabla^\rho\nabla^\sigma(\xi\partial_\rho\varphi\partial_\sigma\varphi)\\
    &\quad -\tfrac{1}{2\sqrt{6}}\Big[\nabla_\mu(\xi\partial\varphi\partial\varphi)\partial_\nu\varphi+\nabla_\nu(\xi\partial\varphi\partial\varphi)\partial_\mu\varphi-g_{\mu\nu}\nabla_\lambda(\xi\partial\varphi\partial\varphi)\partial^\lambda\varphi\Big]\\
    &\quad -\tfrac{1}{3}\xi\Big(2R_{\nu\lambda}\partial_\mu\varphi\partial^\lambda\varphi+2R_{\mu\lambda}\partial_\nu\varphi\partial^\lambda\varphi-g_{\mu\nu}G^{\rho\sigma}\partial_\rho\varphi\partial_\sigma\varphi\\
    &\qquad -R\partial_\mu\varphi\partial_\nu\varphi-R_{\mu\nu}\partial\varphi\partial\varphi-\sqrt{\tfrac{3}{2}}\partial_\mu\varphi\partial_\nu\varphi\Box\varphi\Big)=0,
\end{split}
\end{align}
\twocolumngrid
where $\xi\equiv \xi(\varphi)=4\alpha_\text{GB} e^{\sqrt{2/3}\varphi}$.

Let us also note that the higher-derivative sector falls within the generalized Galileon/Horndeski family
\cite{Horndeski:1974wa,Deffayet:2011gz,Deffayet:2013lga,Kobayashi:2019hrl} (for instance, one can verify that all higher derivatives of $\varphi$ cancel in the equations of motion, leaving second-order dynamics). This explicitly proves that the quadratic corrections in the pre-geometric-L-gravity stemming from the extended MM theory are already dynamically relevant
for cosmological models. In a FLRW background, Eqs.\ \eqref{KG} and \eqref{EFE} highly simplify (see Refs.\ \cite{Addazi:2025qra,Addazi:2025agg}).

Furthermore, such a large value of $\alpha_\text{GB}$ can be translated in terms of the propagation of a nearly massless mode, as $m^2 \sim M^2 \sim \alpha_\text{GB}^{-1} M_\text{P}^2 \sim \Lambda$. This suggests that the quadratic MM model naturally predicts the presence of a scalaron behaving as dynamical dark energy. The parameter $\xi$ can be interpreted as a dynamical degree of freedom that promotes the gravitational $\theta$-angle from a constant to an ultra-light axion-like particle (ALP):
\begin{equation}
\label{constants}
{\theta = 32\pi^2\alpha_{\text{GB}}\longrightarrow \frac{b}{F} = 32\pi^2\alpha_{\text{GB}}\,e^{\sqrt{2/3}\varphi} = 32\pi^2\alpha_{\text{GB}}\,f'} ,
\end{equation}
where $F = M_{\text{P}}$ is a natural identification and $f'$ is related to the dynamical field. 
{Moreover, as has been shown recently, the modulus of this angle is exactly proportional to the Pontryagin theta-angle, once again demonstrating the predictive power of this theory \cite{Addazi:2026eto,Addazi:2026suu}. }

In this interpretation, Eq.\ \eqref{LL} provides the effective Lagrangian governing the ALP \cite{Alexander:2025qkx}. Moreover, non-perturbative quantum gravity effects can generate a periodic effective potential that corrects Eq.~\eqref{LL} and induces an effective super-light mass $m_b$ for the gravi-axion. 
In the $\varphi$ scalaron frame, the gravi-axion mass term is manifest as an exponential correction to the self-interaction potential.
When the Hubble rate becomes comparable to this mass, $m_b \sim H(z_c)$, the axion is effectively unfrozen, which may generate a transient epoch during which the dark-energy equation of state deviates significantly from $w = -1$ through the Gauss--Bonnet coupling and non-minimal kinetic terms. This interpretation is conceptually related to late-time out-of-equilibrium axion models \cite{Montani:2026Axion}. There the axion is a CDM-like component and the effective BGK rate is introduced phenomenologically.

\vspace{0.2cm}

{\it Numerical Results}. We perform the observational analysis in the Jordan frame, which is the matter frame relevant to the pre-geometric construction. 
Here and below, the ``hard-vacuum'' branch denotes the normalized background branch in which the present vacuum density is imposed through
$\Omega_{\Lambda0}=1-\Omega_{m0}-\Omega_{r0}$, while
$\bar\lambda\equiv\lambda/H_0^2$ is fixed by the smaller root of
$\bar\lambda\bigl(1-\bar\lambda/(3\bar M^2)\bigr)=3\Omega_{\Lambda0}$,
with $\bar M\equiv M/H_0$. The scalaron is additionally required to have settled close to the minimum of its potential by the present epoch. Here
$u\equiv d\varphi/dN_E$, with $N_E=\ln a_E$.

The dimensionless parameters used to define the selected hard-vacuum background are
\begin{equation}
\begin{aligned}
 x_{\rm sel}&=\left(\log_{10}\bar M,\Omega_{m0},u_{\rm init},\log_{10}\bar\alpha\right)\\
 &=\left(0.333938,0.298196,0.757921,-3.007676\right).
\end{aligned}
\end{equation}
This corresponds to $\bar M=2.15744$ and $\bar\alpha=9.825\times10^{-4}$. Here $\bar\alpha$ is the effective late-time coefficient entering the numerical background and should not be confused with the bare Planck-unit coefficient $\alpha_{\rm GB}M_{\rm P}^{2}\sim10^{120}$ in Eq.~\eqref{eq:alpha-gb-large}. As established in the earlier matching analysis, the strict minimal relation would give $\bar\alpha_{\rm MM}=1/(8\Omega_{\rm DE,0})\simeq0.178$; the smaller value above is therefore retained as the previously identified effective matching issue and is not claimed to realize the bare MM relation directly. This point was selected by a joint Pareto search over the hard-vacuum parameters and the complexity of a data-independent high-redshift return. The balanced solution moves the visible onset of the phantom excursion from $z\simeq1.07$ to $z\simeq0.90$ while preserving the BAO fit and avoiding a prescribed constant plateau. It is used as a benchmark background, not as a converged posterior estimate. The earlier exploratory \texttt{emcee} stress test used 24 walkers for 300 steps, discarded the first 80 steps and thinned the remainder by a factor of two. The sampled boxes were $0.05\leq\log_{10}\bar M\leq1.20$, $0.24\leq\Omega_{m0}\leq0.36$, $0.25\leq u_{\rm init}\leq1.20$, and $-3.80\leq\log_{10}\bar\alpha\leq-1.70$. A weak Gaussian theory prior with mean $118$ and width $1.25$ was imposed on $\log_{10}(\alpha_{\rm GB}M_{\rm P}^{2})$, while no strict gravitational-wave prior was imposed during sampling. The mean acceptance fraction was $0.122$, and the split-$\widehat R$ values ranged from $2.25$ to $2.99$; consequently, the chain is used only as a diagnostic and no posterior intervals or corner-plot constraints are quoted.

\paragraph{Data and likelihood convention.} We use the 13-element DESI DR2 BAO distance vector spanning $0.295\leq z\leq2.33$ \cite{DESI:DR2BAO}. For growth we use an auxiliary six-element compressed $f\sigma_8$ vector at $z=0.295,0.510,0.706,0.934,1.321,$ and $1.484$, employing the redshift bins of the DESI DR1 full-shape analysis \cite{DESI:DR1FS}. Ordered by increasing redshift, the vector is
\begin{equation}
\begin{split}
\bm d_{\rm FS}=(&0.4543526,0.4563152,0.4539623,\\
&0.4444241,0.4256204,0.3585684).
\end{split}
\end{equation}
This vector and its $6\times6$ covariance are used only as a low-dimensional diagnostic; they are neither the native nor an official standalone DESI full-shape likelihood. The BAO and growth covariances are available separately, but a matched BAO--FS cross-covariance is not available for this mixed data-vector combination. We therefore report $\chi^2_{\rm BAO}$ and $\chi^2_{\rm FS}$ separately. Their block-diagonal sum is quoted only as an auxiliary diagnostic and is not used for formal model selection.

For each background we analytically profile the BAO amplitude
\begin{equation}
 S\equiv\frac{c}{H_0r_d},
\end{equation}
and the growth amplitude $A\equiv\sigma_8(0)$. The Jordan-frame BAO templates are
\begin{align}
\frac{D_M(z)}{r_d}&=S\int_0^z\frac{dz'}{E_J(z')},\\
\frac{D_H(z)}{r_d}&=\frac{S}{E_J(z)},\\
\frac{D_V(z)}{r_d}&=S\left[\frac{z}{E_J(z)}\left(\int_0^z\frac{dz'}{E_J(z')}\right)^2\right]^{1/3}.
\end{align}
The growth calculation is evaluated with CAMB at the exact six redshifts, and the shape $fD(z)=f\sigma_8(z)/\sigma_8(0)$ is multiplied by the profiled amplitude $A$.

\paragraph{Perturbative audit of the covariant action.} A direct $\mathcal H$-EFTCAMB implementation of the Horndeski functions associated with Eq.~\eqref{L_canonical} was previously shown to reproduce the hard-vacuum background to better than $2\times10^{-4}$ over $0\leq z\leq30$ \cite{HEFTCAMB}. Across the normalized hard-vacuum branches examined in the exact mapping, $c_T^2$ becomes negative at intermediate redshift and the low-redshift deviation from luminality remains many orders of magnitude above the multimessenger bound of order $10^{-15}$ \cite{GW170817GRB}. Scans in $\bar\alpha$ and $u_{\rm init}$ did not identify a normalized, vacuum-settled branch satisfying both tensor stability and the low-redshift speed constraint. The newly selected balanced branch is therefore used only as the background input to the massive gravi-axion framework; the unmodified covariant action is not used as the final perturbation theory.

\begin{figure}[t]
\centering
\includegraphics[width=\columnwidth]{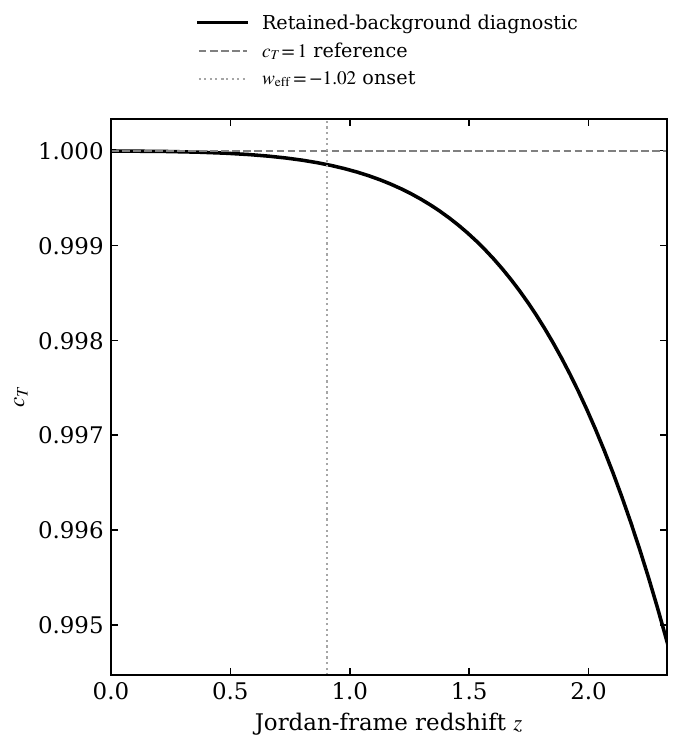}
\begingroup

\caption{Tensor speed on the retained numerical trajectory for $0\leq z\leq2.33$, (evaluated with the positive effective
Gauss--Bonnet coupling and the $\alpha_{\rm GB}/2$ normalization).
The horizontal dashed line marks $c_T=1$; the vertical dotted line marks
the $w_{\rm eff}=-1.02$ onset threshold at $z\simeq0.903$. }
\label{fig:ct_diagnostic}
\endgroup
\end{figure}

\begingroup

Figure~\ref*{fig:ct_diagnostic} shows an off-shell tensor-speed diagnostic
on the retained  numerical trajectory, with the positive effective
Gauss--Bonnet coupling and the $\alpha_{\rm GB}/2$ normalization of
Eq.~(\ref*{L_canonical}). For the full operator combination in that action,
the tensor speed in reduced Planck units is
\begin{equation}
\label{eq:ct_diagnostic}
c_T^2=
\frac{1+4\alpha_{\rm GB}e^{k\varphi}
\bigl(k\ddot\varphi+\dot\varphi^2\bigr)}
{1+4\alpha_{\rm GB}e^{k\varphi}
\bigl(kH\dot\varphi-\dot\varphi^2/3\bigr)},
\qquad k=\sqrt{2/3}.
\end{equation}
Here $H$ and the time derivatives are evaluated in the Einstein frame,
while the plotted redshift is the Jordan-frame redshift. The expression
includes the contributions of both the scalar-dependent Gauss--Bonnet term
and the non-minimal derivative coupling.

Over $0\leq z\leq z_{\rm join}=2.33$, the speed increases toward unity
as the redshift decreases, with $c_T(0)-1\simeq-3.44\times10^{-6}$.
The absolute deviation therefore remains above the multimessenger bound.
The vertical marker identifies the $w_{\rm eff}=-1.02$ onset threshold
at $z\simeq0.903$. The actual $w_{\rm eff}=-1$ crossing occurs at
$z\simeq2.856$ in the effective-density return described below, outside
the plotted interval. 
\par
\endgroup

\paragraph{Effective gravi-axion framework.} To calculate linear observables without modifying the microscopic pre-geometric action, we organize the late-time sector around a vacuum-like gravi-axion response acting on $y_{\rm DE}=\ln(\rho_{\rm DE}/\rho_{c0})$. The minimal local response is
\begin{equation}
\begin{aligned}
 W(X)&=\frac{(1+p_{\rm low}/p_{\rm high})X^{p_{\rm low}}}
 {1+(p_{\rm low}/p_{\rm high})X^{p_{\rm low}+p_{\rm high}}},\\
 X(z)&=\frac{E_{\rm hv}(z)}{E_{\rm hv}(z_c)},\\
 y_{\rm DE}(z)&=\ln\Omega_{\Lambda0}-A_b\,W[X(z)],
\end{aligned}
\end{equation}
where $E_{\rm hv}$ denotes the selected hard-vacuum expansion. This response peaks at $H(z_c)\simeq m_b$ and is suppressed in both asymptotic regimes. The selected parameters are $A_b=0.78283$, $z_c=2.91396$, $p_{\rm low}=5.22721$ and $p_{\rm high}=5.06765$, corresponding to $m_b/H_0=4.22661$ or $m_b\simeq6.08\times10^{-33}\,{\rm eV}$. The visible $w_{\rm eff}<-1.02$ onset occurs at $z\simeq0.86$. Relative to the smooth numerical background representative used below, the response has an rms displacement of $4.6\times10^{-4}$ in $E(z)$ over $0\leq z\leq2.33$ and $4.5\times10^{-4}$ in the unscaled 13-entry BAO distance vector, while $\rho_{\rm DE}$ remains positive and the response becomes negligible at high redshift. The latter statistic is a shape diagnostic, not a new covariance $\chi^2$. A one-pole causal-memory extension improves the full-range expansion rms by only about $15\%$ and places $z_c$ at the boundary of the adopted interval, so the local response is retained.

For the full CAMB calculation, we use a constrained smooth background representative of this response class. It follows the hard-vacuum background through the highest BAO redshift, $z_{\rm join}=2.33$, to numerical interpolation accuracy, then returns smoothly to the same-parameter GR/$\Lambda$CDM background and becomes exactly $\Lambda$CDM at $z_{\rm end}=5.15$. The high-redshift return is constructed in $y(x)=\ln[\rho_{\rm DE}(x)/\rho_{c0}]$, with $x=\ln(1+z)$, as a constrained cubic B-spline minimizing $\int [d^2y/dx^2]^2dx$. It matches $y$ and $dy/dx$ to the hard-vacuum solution at $z_{\rm join}$, reaches the same-parameter $\Lambda$CDM density with $dy/dx=0$ at $z_{\rm end}$, maintains a positive density and $w_{\rm eff}\leq0$, and permits only one smooth rebound. No constant plateau is prescribed. The local response reproduces the expansion of this representative but not its exact density endpoint: at $z_{\rm end}$ the density differs by about $6.9\%$, while the corresponding relative displacement in $E$ is only $3.4\times10^{-4}$. The spline should therefore not be interpreted as an exact gravi-axion field solution. The tensor and local gravitational sectors tend to be  GR-like, with $c_T=1$, and the dark-energy perturbations are evolved with the PPF prescription \cite{Fang:2008sn}. For reproducibility, the CAMB calculation fixes $H_0=67.4~{\rm km\,s^{-1}\,Mpc^{-1}}$, $\Omega_{m0}=0.298196$, $\Omega_{r0}=9.1923\times10^{-5}$, $\omega_b\equiv\Omega_bh^2=0.02237$, $A_s=2.1\times10^{-9}$, $n_s=0.965$, $\tau=0.054$, $N_{\rm eff}=3.044$, and $\sum m_\nu=0$, with the PPF rest-frame sound speed fixed to $c_s^2=1$.

\begin{figure*}[t]
\centering
\includegraphics[width=0.48\textwidth]{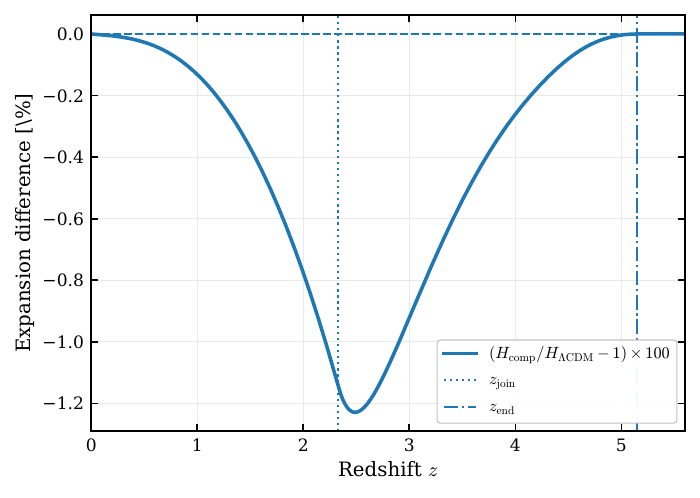}\hfill
\includegraphics[width=0.48\textwidth]{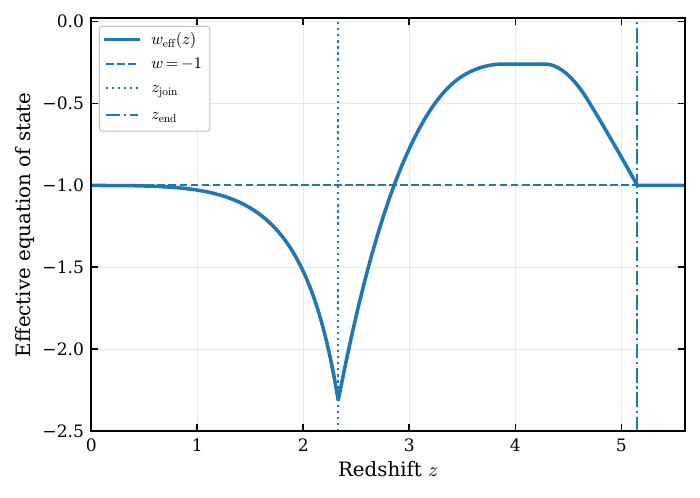}
\caption{Gravi-axion benchmark background. Left: fractional difference between the effective Jordan-frame expansion and same-parameter $\Lambda$CDM. The smooth benchmark follows the hard-vacuum solution through $z_{\rm join}=2.33$ and becomes exactly $\Lambda$CDM at $z_{\rm end}=5.15$. Right: reconstructed effective equation of state, showing a transient crossing of the phantom divide.}
\label{fig:background_completion}
\end{figure*}

The effective density stays positive throughout the smooth return, with $\rho_{\rm DE}^{\rm min}/\rho_{c0}=0.334$, while the reconstructed equation of state spans
\begin{equation}
 -2.31\lesssim w_{\rm eff}(z)\lesssim-0.26.
\end{equation}
The hard-vacuum branch reaches the visible phantom-onset threshold near $z\simeq0.90$, after which the constrained high-redshift rebound returns the benchmark to $\Lambda$CDM. Thus the effective gravi-axion framework exhibits a transient effective phantom crossing, of the type considered in recent DESI dynamical-dark-energy analyses \cite{DESI:ExtendedDE}. This crossing refers to the effective Jordan-frame fluid in a non-equilibrium phasex and is not, by itself, evidence for a fundamental negative-kinetic-energy scalar. The background and $w_{\rm eff}$ evolution are shown in Fig.~\ref{fig:background_completion}.

\begin{figure*}[t]
\centering
\includegraphics[width=0.98\textwidth]{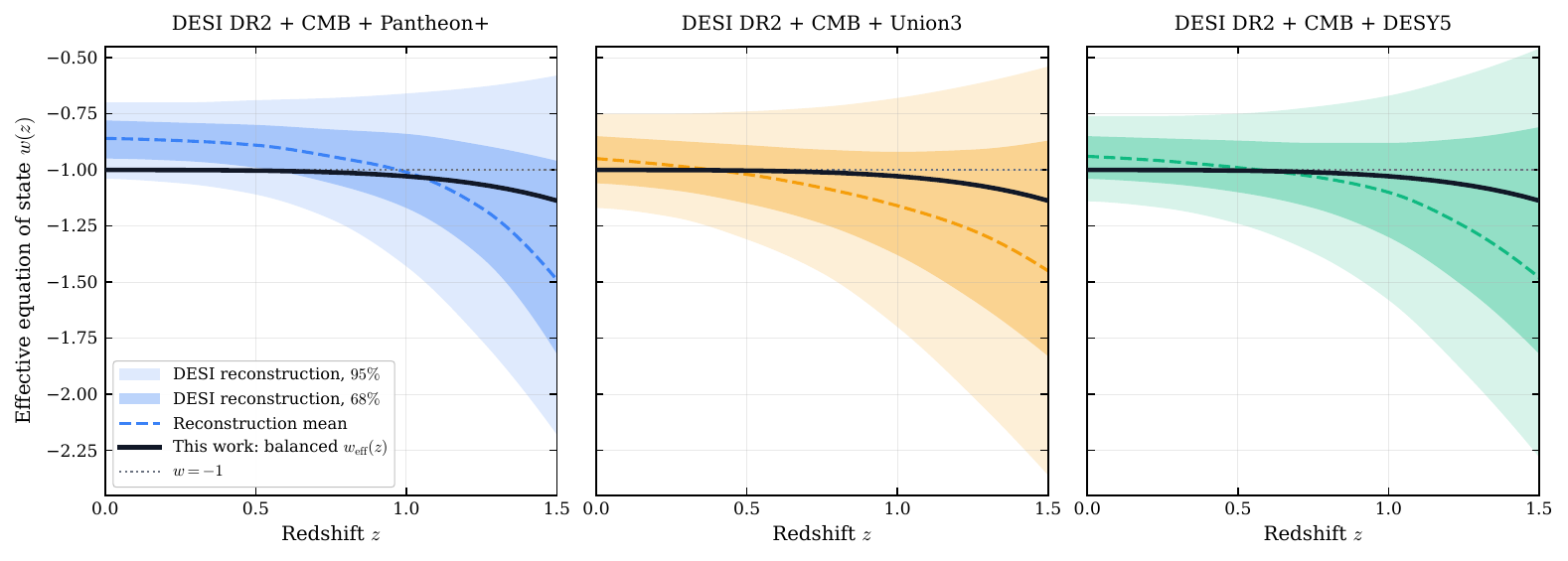}
\caption{Low-redshift effective equation of state of the gravi-axion benchmark compared with observational reconstructions. The black solid curve shows $w_{\rm eff}(z)$ for the benchmark. The coloured dashed curves and shaded regions show the reconstructed means and the approximate $68\%$ and $95\%$ confidence regions obtained from DESI DR2 BAO and CMB combined, from left to right, with Pantheon+, Union3 and DESY5 supernovae. The horizontal dotted line denotes the cosmological-constant value $w=-1$. The bands are from Ref.~\cite{Yang:2025QuintomMG}, as reproduced and discussed in Ref.~\cite{Cai:2026QuintomReview}.}
\label{fig:weff_desi_bands}
\end{figure*}

To place this evolution in observational context, Fig.~\ref{fig:weff_desi_bands} compares the effective equation of state of the gravi-axion benchmark with reconstructed confidence regions obtained from DESI DR2 BAO, CMB and the Pantheon+, Union3 and DESY5 supernova samples \cite{Yang:2025QuintomMG,Cai:2026QuintomReview,Brout:2022PantheonPlus,Rubin:2025Union3,DES:2024Y5}. The benchmark remains close to $w=-1$ at low redshift and evolves toward more negative values as the redshift increases. In this approximate visual comparison, the model curve remains inside the displayed $95\%$ regions throughout $0\leq z\leq1.5$ and overlaps substantial parts of the $68\%$ regions.

\begin{figure*}[t]
\centering
\includegraphics[width=0.48\textwidth]{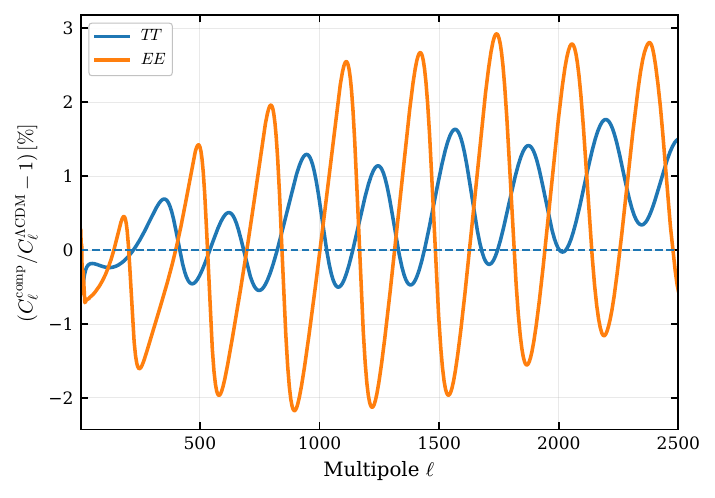}\hfill
\includegraphics[width=0.48\textwidth]{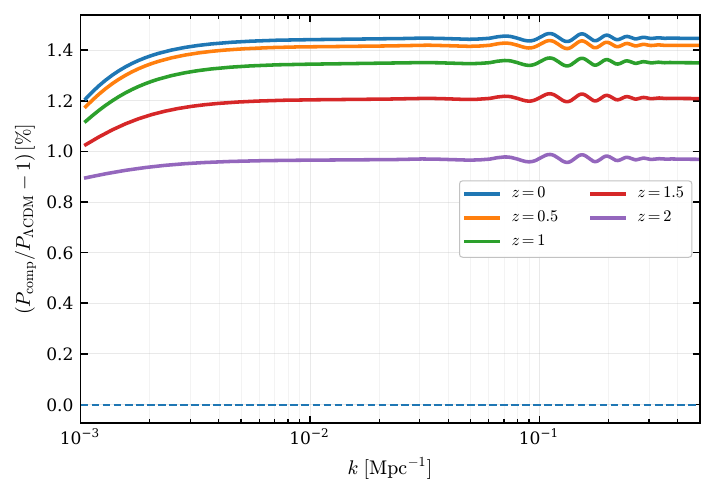}
\caption{Full linear Boltzmann observables for the gravi-axion benchmark with the imposed GR+PPF prescription, relative to the same-parameter $\Lambda$CDM model. Left: fractional changes in the CMB $TT$ and $EE$ spectra. Right: fractional changes in the linear matter power spectrum at representative redshifts.}
\label{fig:boltzmann_ratios}
\end{figure*}

\paragraph{Full Boltzmann evolution and low-redshift likelihoods.} The gravi-axion benchmark, with the GR+PPF prescription specified above, was evolved with CAMB from the radiation era to the present. All CMB, lensing, transfer-function and matter-power outputs remain finite. The converged present-day values are
\begin{equation}
 \sigma_8(0)=0.79849,\qquad f\sigma_8(0)=0.40818,
\end{equation}
with a $k_{\rm max}$ convergence error of $1.8\times10^{-5}$. Relative to the same-parameter $\Lambda$CDM model, the largest changes are approximately $1.77\%$ in $C_\ell^{TT}$, $2.92\%$ in $C_\ell^{EE}$ and $1.47\%$ in the linear matter spectrum over the plotted range. These ratios are displayed in Fig.~\ref{fig:boltzmann_ratios}.

The separate low-redshift likelihood contributions are summarized in Table~\ref{tab:likelihoods}. The gravi-axion benchmark is slightly worse in BAO and remains close to the baseline in the growth sector. With profiled $S$ and $A$, the auxiliary block-diagonal difference is $\Delta\chi^2=0.459$ relative to the common-data flat-$\Lambda$CDM baseline. The fixed-$A_s$ growth diagnostic gives $\chi^2_{\rm FS}=5.91$ for six measurements, showing that the agreement is not solely produced by profiling the amplitude.

\begin{table}[t]
\caption{Separate BAO and auxiliary compressed-growth likelihoods. The last column is the block-diagonal sum and is quoted only as an auxiliary diagnostic.}
\label{tab:likelihoods}
\centering
\begin{tabular}{lccc}
\toprule
Model & $\chi^2_{\rm BAO}$ & $\chi^2_{\rm FS}$ & $\chi^2_{\rm block}$\\
\midrule
Flat $\Lambda$CDM & 10.3037 & 5.6734 & 15.9770\\
Gravi-axion & 10.7118 & 5.7241 & 16.4359\\
Difference & $+0.4081$ & $+0.0507$ & $+0.4588$\\
\bottomrule
\end{tabular}
\end{table}

\begin{figure*}[t]
\centering
\includegraphics[width=0.32\textwidth]{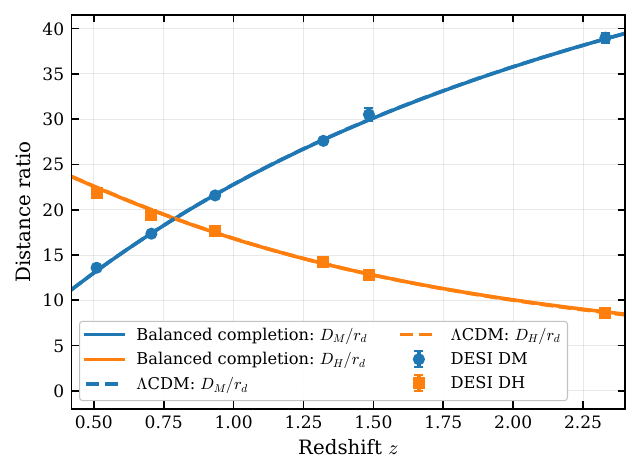}\hfill
\includegraphics[width=0.32\textwidth]{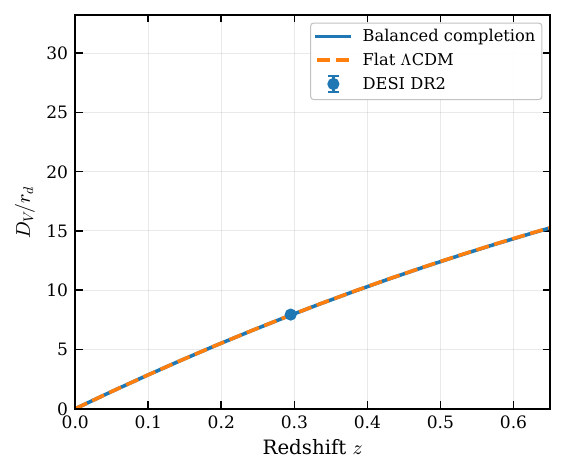}\hfill
\includegraphics[width=0.32\textwidth]{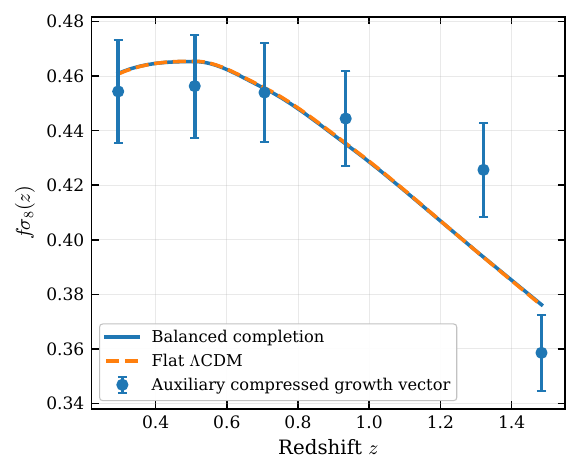}
\caption{DESI BAO measurements, the auxiliary compressed growth diagnostic and the Jordan-frame gravi-axion benchmark. Left: anisotropic BAO observables $D_M/r_d$ and $D_H/r_d$. Middle: the isotropic $D_V/r_d$ measurement. Right: the auxiliary compressed growth vector $f\sigma_8(z)$. Solid curves show the gravi-axion benchmark and dashed curves the common-data flat-$\Lambda$CDM baseline. Error bars display the diagonal uncertainties; the full within-sector BAO and FS covariance matrices are used separately in the likelihood.}
\label{fig:desi_completion}
\end{figure*}

The BAO and growth comparison is shown in Fig.~\ref{fig:desi_completion}. The resulting benchmark is statistically close to $\Lambda$CDM on these low-redshift data, but it should be interpreted as a fixed-point viability audit rather than a completed posterior analysis. In particular, we do not assign a reduced $\chi^2$, AIC or BIC to the framework, since its effective response and high-redshift representative have not been incorporated into a consistently parameter-counted global fit.

\twocolumngrid





\vspace{0.2cm}

{\it Conclusion}. In this work, we have demonstrated that a quadratic extension of MacDowell--Mansouri pre-geometric gravity renders the Gauss--Bonnet $\theta$-term dynamically relevant following spontaneous symmetry breaking. This framework naturally gives rise to a scalar-tensor theory featuring an ultra-light degree of freedom in the hard-vacuum sector. The interconnection between the bare Gauss--Bonnet coefficient, the cosmological constant, and the de Sitter entropy constitutes a central theoretical pillar of our construction, with the quadratic model providing a concrete foundation for dynamical dark energy.

Our numerical analysis yields two logically distinct results. First, the Jordan-frame hard-vacuum expansion history successfully reproduces the DESI DR2 BAO data, and a transient gravi-axion response localized near $H\simeq m_b$ reproduces the benchmark expansion and BAO-distance shape while becoming negligible at high redshift. Second, the perturbation sector of the unmodified covariant action proves non-viable across the full cosmological timeline: its exact Horndeski mapping exhibits a tensor-gradient instability and fails to satisfy the multimessenger speed constraint. Holographic or topological arguments regarding the radiative stability of the hierarchy cannot compensate for these kinetic and gradient shortcomings.
However, we have shown that this situation occurs only in the physically implausible limit of a perfectly massless gravi-axion. We therefore relax this assumption by introducing a natural dark-energy-scale mass for the gravi-axion, which corresponds to a new exponential term in the dual scalaron field $\varphi$.

To implement this gravi-axion framework in the linear regime, we adopt a constrained smooth background representative that follows the hard-vacuum Jordan-frame expansion up to $z=2.33$ and transitions back to GR/$\Lambda$CDM by $z=5.15$. The tensor and local sectors are taken to be GR-like, while the effective dark-energy perturbations are evolved using the PPF formalism. The resulting benchmark exhibits a transient effective phantom crossing, maintains a positive effective energy density, and admits a fully regular CAMB evolution. Its separate likelihoods---$\chi^2_{\rm BAO}=10.712$ and $\chi^2_{\rm FS}=5.724$---differ only marginally from the corresponding $\Lambda$CDM values of $10.304$ and $5.673$ obtained from the same datasets.

The pre-geometric theory thus provides a well-motivated dynamical origin for the background and an ultra-light gravi-axion degree of freedom, while the effective response framework offers a concrete description of its late-time phenomenology. Standard misalignment supplies the qualitative mechanism: at high redshift, $H\gg m_b$ freezes the field through Hubble friction, while its dynamics become relevant near $H\sim m_b$. The chosen positive-density response is centered at $z_c\simeq2.91$, reproduces the benchmark expansion and BAO-distance shape to better than $10^{-3}$, and decays at high redshift. This establishes background-level compatibility between the gravi-axion framework and the fitted hard-vacuum evolution. The construction therefore provides a concrete phenomenological link between pre-geometric gauge gravity, gravi-axion dynamics, and cosmic acceleration, while offering a solid foundation for further investigation into the full perturbative regime.

\vspace{0.2cm}

{\bf Acknowledgments.}
AA's work is supported by the {\it Program for Innovative Research Team} in Anqing Normal University. YC is supported in part by NSFC (12433002), by CAS young interdisciplinary innovation team (JCTD-2022-20), by 111 Project (B23042), by CSC Innovation Talent Funds, and by USTC Fellowship for International Cooperation. SC and GM acknowledge the support of Istituto Nazionale di Fisica Nucleare (INFN), Sez.\ di Napoli, {\it Iniziative Specifiche} QGSKY and MoonLight-2. SC is grateful to the {\it Gruppo Nazionale di Fisica Matematica} (GNFM) of {\it Istituto Nazionale di Alta Matematica} (INDAM) for support. This publication is based upon work from COST Action CA21136 -- ``Addressing observational tensions in cosmology with systematics and fundamental physics (CosmoVerse)'', supported by COST (European Cooperation in Science and Technology).

\end{document}